# KOLMOGOROV COMPLEXITY SPECTRUM FOR USE IN ANALYSIS OF UV-B RADIATION TIME SERIES

DRAGUTIN T. MIHAILOVIĆ

*Faculty of Agriculture, University of Novi Sad,*

*Dositeja Obradovica Sq. 8, 21000 Novi Sad, Serbia*

*guto@polj.uns.ac.rs*

SLAVICA MALINOVIĆ-MILIĆEVIĆ

*ACIMSI - University Center for Meteorology and Environmental Modeling, University of Novi Sad,*

*Dositeja Obradovica Sq. 3, 21000 Novi Sad, Serbia*

ILIJA ARSENIĆ

*Faculty of Agriculture, University of Novi Sad,*

*Dositeja Obradovica Sq. 8, 21000 Novi Sad, Serbia*

*guto@polj.uns.ac.rs*

NUSRET DREŠKOVIĆ

*Faculty of Sciences, Department of Geography, University of Sarajevo,*

*Zmaj from Bosnia 33-35, 71000 Sarajevo, Bosnia and Herzegovina*

BEATA BUKOSA

*Faculty of Sciences, Department of Physics, Center for Meteorology and Environmental Predictions, University of Novi Sad,*

*Dositeja Obradovica Sq. 3, 21000 Novi Sad, Serbia*

ABSTRACT

We have used the Kolmogorov complexity and sample entropy measures to estimate the complexity of the UV-B radiation time series in the Vojvodina region (Serbia) for the period 1990–2007. We have defined the Kolmogorov complexity spectrum and have introduced the Kolmogorov complexity spectrum highest value (KLM). We have established the UV-B radiation time series on the basis of their daily sum (dose) for seven representative places in this region using: (i) measured data, (ii) data calculated via a derived empirical formula and (iii) data obtained by a parametric UV radiation model. We have calculated the Kolmogorov complexity (KL) based on the Lempel-Ziv Algorithm (LZA), KLM and Sample Entropy (SE) values for each time series. We have divided the period 1990–2007 into two subintervals: (a) 1990-1998 and (b) 1999-2007 and calculated the KL, KLM and SE values for the various time series in these subintervals. It is found that during the period 1999-2007, there is a decrease in the KL, KLM, and SE, comparing to the period 1990–1998. This complexity loss may be attributed to (i) the increased human intervention in the post civil war period causing increase of the air pollution and (ii) the increased cloudiness due to climate changes.

## 1. INTRODUCTION

### *1.1 Preliminaries*

The notion of complexity is defined on different ways, but without definition of complexity, which is clearly established in the scientific literature. If we follow Grassberger [1], on the level of intuition, it can be accepted as something that is placed in between uniformity and total randomness. When we use the term complexity in physical systems we explicitly think that it is a measure of the probability of the state vector of the system. It is a mathematical measure, one in which two distinct states are never be combined into a composite whole and considered equal, as is done for the notion of entropy in statistical mechanics. The physical complexity of a sequence "refers to the amount of information that is stored in that sequence about a particular environment".[2] This should not be confused with mathematical (Kolmogorov) complexity; it is a distinct mathematical complexity, which only deals either with the intrinsic regularity or irregularity of a sequence in this case.

The quantification of the complexity of a system is one of the aims of non-linear time series analysis. Complexity of the system is hidden in the dynamics of the system. However, if there is no recognizable structure in the system, it is considered to be stochastic. Because of the occurrence of noise, spurious experimental result and artifacts in various forms, it is often not easy to get reliable information from a series of measurements. The time series is only information about physical state, we obtained by the measurement, and accordingly that is only source for establishing the level of physical complexity. Our integrated knowledge about different parts of the physical system usually comes from either measurements or physical models having different levels of sophistication [3]. Therefore, analysis of the modeled or measured time series gives deeper insights into a physical phenomenon.

*1.2. Complexity of the UV-B radiation time series*

Influenced by climate, vegetation, geography and human factors, many meteorological elements as well as the UV-B radiation in a specific geographic region may range from being relatively simple to complex, which exhibits significant variability in both time and space. Recently, the human factor becomes the most important issue regarding the complexity of the meteorological elements. Namely, actions in the form of different human activities in environment (air, soil and water) can be either constructive or destructive. They (i) can have positive or negative impact on the human economy and (ii) can leave landscape features that are present for a long time. Thus, it is of interest to determine the nature of complexity in the UV-B radiation processes that can not be done by traditional methods. This approach requires the use of various measures of complexity to get a deeper insight into the complexity of the UV-B radiation, which may provide us: (i) more comprehensive investigation of possible change in UV-B radiation due to human activities and response to climate change and (ii) improving the application of the stochastic process concept in radiation its modeling, forecasting, measuring and other ancillary purposes. [4-9]

Kolmogorov Complexity (KL in further text) is a measure of descriptive complexity contained in an object. As defined above it refers to the minimum length of a program such that a universal computer can generate a specific sequence. Following Kolmogorov's idea, Lempel & Ziv [10] developed an algorithm for computing this complexity based on symbolic dynamics. This is a non parametric measure of complexity for finite sequences and is related to the number of distinct substrings and the rate of their occurrences. It is a measure of the disorder or irregularity in a sequence. It has been used for the analysis of complexity of time series in a wide variety of applications ranging from biological and biomedical systems, geophysical, environmental to financial markets.[11-13] Entropy is commonly used to characterize the complexity of a time series also including radiation ones. Thus, approximate entropy with a biased statistic, is effective for analyzing the complexity of noisy, medium-sized time series. [14] Richman & Moorman [15] proposed another statistics, sample entropy (SE), which is unbiased and less dependent on data. Traditional entropies quantify only the regularity of time series having some disadvantages. [16] The KL measure has not been used for analyzing the complexity of meteorological time series, while the SE is rarely used for analysis, particularly in investigating the climate complexity. Therefore, it is of interest to investigate how these measures can be employed in complexity analysis of the UV-B radiation dose time series for different purposes.

*1.3. Purpose of the paper*

The purpose of this paper is to investigate the complexity of the UV-B radiation dose time series for places spatially distributed over some area, using the Kolmogorov complexity (KL) and sample entropy (SE) measures. To reinforce this analysis we introduce additional complexity measures based on the Kolmogorov complexity, i.e. the Kolmogorov complexity spectrum and the Kolmogorov complexity spectrum highest value (KLM).

For our analysis we use the Vojvodina region (Serbia). UV-B radiation records in the Vojvodina region (Serbia) are of relatively short size. In order to create the UV radiation time series for seven representative places we have included: (i) values measured in Novi Sad using the broadband Yankee UVB-1 biometer, (ii) values computed by a parametric numerical model and (iii) values calculated by an empirical formula derived on the basis of the linear correlation between the daily sum of the UV-B radiation and the daily sum of the global solar radiation. In the further development we analyze the complexity of the UV-B radiation dose time series from the seven representative places in the Vojvodina region (Serbia) for the period 1990–2007, using the KL, KLM and SE measures. We also investigate the effect of different human activities, events and climate change on the UV-B radiation dose complexity by dividing the period 1990–2007 into two equal subintervals: (a) 1990-1998 and (b) 1999-2007. Namely, there was a evident increase in human activity in the Vojvodina region after 1998 (post civil war period, military activities in air, intensification of economic activity, more intensive traffic, traditional home heating).[17] It has caused high air pollution and further changes in the UV-B radiation dose complexity in the Vojvodina region. The KL and SE values are calculated for the various time series in each of the above subintervals. It is found that during the period 1999- 2007, there is a decrease in complexity in most of the places in comparison to the period 1990–1998. This complexity loss may be attributed to (i) human intervention in the post civil war period that cause larger air pollution and (ii) increased cloudiness due to climate changes. In Section 2: (i) we shortly describe the KL, introducing the KL spectrum and its highest value (2.1), (ii) we elaborate physical background of the UV radiation parametric numerical model (2.2) and (iii) we give the feature of the methodology used in the paper (2.3). In Section 3 we show results, which include statistical evaluation and discussion. Concluding remarks are given in Section 4.

## 2. METHOD AND MATERIAL

### 2.1. The Kolmogorov complexity spectrum and its highest value

Kolmogorov complexity is a measure often use in analysis of physical time series which is obtained either in the process of measurement or from some model. A good introduction to the KL complexity can be found in Ref. 18 and with a comprehensive description in Ref. 19. On the basis of Kolmogorov's idea, Lempel & Ziv [10] developed an algorithm (refer as LZA), which is used in evaluation of randomness of time series as a measure of its disorder. Let us note that the KL complexity is not able to distinguish between time series, which have different amplitude oscillations but very similar random components. A short description of the LZA is given in the Appendix A. However, if we use the measure obtained by the LZA we get just one information about complexity of a time series. This information strongly depends on the choice of the threshold $x_*$. Therefore, it must be chosen from some physical reason, although the threshold is often derived statistically from the analyzed time series. As a consequence of such choice, in both cases, much information about time series could be lost. From that reason we introduce two more complexity measures based on the KL complexity.

First, we create the normalized time series $\{x_i\}$, $i = 1, 2, 3, 4, ..., N$ by the transformation $x_i = (X_i - X_{min})/(X_{max} - X_{min})$, where $\{X_i\}$ is a time series obtained either by a measuring procedure or from a physical model, where $X_{max} = max\{X_i\}$ and $X_{min} = min\{X_i\}$. Then, we make a transformation into a finite symbol string by comparison with series of thresholds $\{x_{t,i}\}$, $i = 1, 2, 3, 4, ..., N$, where each element is equal to the corresponding element in the considered time series $\{x_i\}$, $i = 1, 2, 3, 4, ..., N$. The original signal samples are converted into a 0–1 sequences $\{S_i^{(k)}\}$, $i = 1, 2, 3, 4, ..., N$, $k = 1, 2, 3, 4, ..., N$ defined by comparison with a threshold $x_{t,i}$,

$$S_i^{(k)} = \begin{cases} 0 & x_i < x_{t,k} \\ 1 & x_i \geq x_{t,k} \end{cases}. \quad (1)$$

After we apply the LZA on each element of series $\{S_i^k\}$ (Appendix A) we get the *Kolmogorov complexity spectrum* $\{K_i^C\}$, $i=1,2,3,4,...,N$. This spectrum we introduce to explicitly describe the time series complexity of each element in a time series that contribute to the physical process at whole from which the physical time series (in our case the UV-B radiation dose) comes from. The highest value in this series, i.e. $\max\{K_i^C\}$ we call the *Kolmogorov complexity spectrum highest value* (refer as KLM). Let us note that the KL means that it is computed using the threshold as in Appendix A.

In order to demonstrate the meaning of the measures we have introduced, a time series $\{x_i\}$, $i=1,2,3,4,...,1000$ was generated by a generalized logistic map.[20] Mathematically, that map is written as

$$\Phi(x) = rx^p(1-x^p), \qquad (2)$$

where $r$ is a logistic parameter, $0 < r \leq 4$, which for $p=1$ becomes well known logistic equation. This map expressed the exchange of biochemical substance between cells that is defined by a diffusion-like manner,[21] where the parameter $p$ is the cell affinity ($0 < p \leq 1$). We have chosen this map because it is suitable for an illustration of the meaning the KLM.

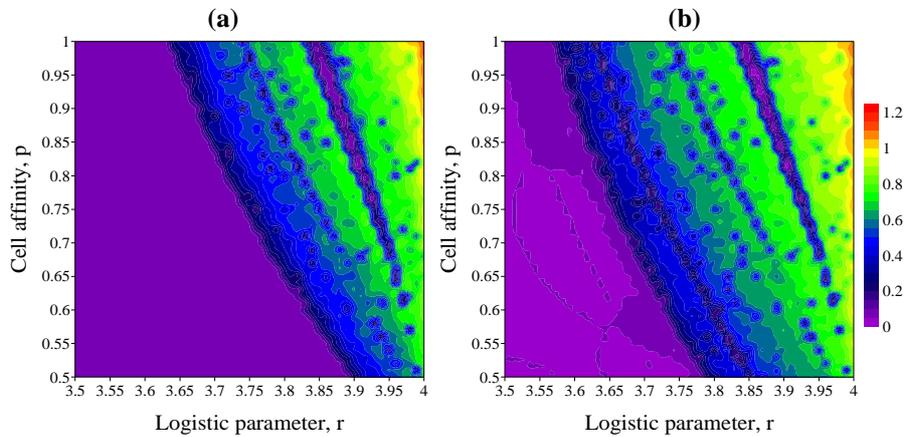

Fig. 1. The dependence on the logistic parameter $r$ and cell affinity $p$ of (a) the Kolmogorov complexity (KL) and (b) the Kolmogorov complexity spectrum highest value (KLM), simulated by the generalized logistic equation $\Phi(x) = ax^p(1-x^p)$.

In order to explore the dependence of the KL and KLM and on (a) the logistic parameter $r$ and (b) the cell affinity $p$ simulated by the generalized logistic Eq. (2) we have performed corresponding computations. In those computations, for each $r$ from 3.5 to 4.0 and $p$ from 0 to 1, with step 0.01, $10^3$ iterations were applied for an initial state, and then the first $10^2$ steps were abandoned. Looking at Figs. 1a-1b, which depicts the KL and KLM complexities, respectively, we can see regions with different levels of complexity. Further inspection of figures point out that in the region of the KLM (Fig. 1b) its values are higher than the KLM ones (Fig. 1a). Apparently, the KLM is better complexity measure for a time series than the KL one. This is because the KL carries average information about the time series. In contrast to that the KLM carries the information about the highest complexity among all complexities in the Kolmogorov complexity spectrum. Therefore, this measure should be included in the complexity analysis of time series of different origin since it gives deeper insights into their complexities.

### 2.2. Short physical background of the UV radiation parametric numerical model

We have partly generated time series of the UV-B radiation by a parametric numerical model NEOPLANTA. This model computes the solar direct and diffuse UV irradiances on a horizontal surface under cloud free conditions for the wavelength range 280–400 nm with 1-nm resolution as well as the UVI. Model simulates the effects of the absorption of the UV radiation by ozone ($O_3$), sulphur dioxide ($SO_2$) and nitrogen dioxide ($NO_2$) and absorption and scattering by aerosol and air molecules in the atmosphere. Atmosphere in model is divided in 40 parallel layers with constant values of meteorological parameters. Its vertical resolution of the model is 1 km for altitudes less than 25 km and above this height 5 kilometres layers were employed. The required input parameters are: the local geographic coordinates and time or solar zenith angle, altitude, spectral albedo, and the total amount of gases. The NEOPLANTA model includes its own vertical gas profiles [22] and extinction cross sections [23,24], extraterrestrial solar irradiance shifted to terrestrial wavelength [25], aerosol optical properties for 10 different aerosol types [26], and spectral albedo for nine different ground surface types [22]. The model uses standard atmosphere meteorological profiles although it is possible to include assimilation of real time meteorological data assimilated from the high level resolution atmospheric mesoscale models. Output data are spectral direct, diffuse, and global irradiance divided into the UV-A (320– 400 nm) and UV-B (280– 320 nm) part of the spectrum, erythemally weighted UV irradiance calculated using the erythemal action spectrum by McKinley & Diffey [27], the UVI, spectral optical depth, and spectral transmittance for each atmospheric component. All outputs are computed at the lower boundary of each layer.

The UV irradiance is calculated as the sum of the direct and the diffuse components. Calculation of the direct part of radiation is carried out by the Beer-Lambert law. The direct irradiance $I_{dir}(\lambda)$ at wavelength $\lambda$ received at ground level by unit area is given by

$$I_{dir}(\lambda) = I_0(\lambda) T(\lambda), \tag{3}$$

where $I_0(\lambda)$ is the extraterrestrial irradiance corrected for the actual Sun-Earth distance and $T(\lambda)$ is the total transmittance that includes $O_3$, $SO_2$, $NO_2$, aerosol and air transmittances. Each of individual transmittances is calculated using optical depth $\tau(\lambda)$ that is the product of extinction coefficient $\beta(\lambda)$ and ray path through the atmosphere $s$

$$T(\lambda) = exp(-\tau(\lambda)) = exp(-\beta(\lambda)s). \tag{4}$$

Extinction coefficient of UV radiation $\beta$ is calculated by the product of the cross-sectional area $\sigma$ and layer particle concentration $N$

$$\beta(\lambda) = \sigma(\lambda) N. \tag{5}$$

The starting point for calculation of diffuse part of radiation is the set of equations from Bird and Riordan spectral model [28], which represents equations from previous parametric models [29, 30], improved after comparisons with rigorous radiative transfer model and with measured spectra. The diffuse irradiance $I_{dif}(\lambda)$ is divided into three components: (i) the Rayleigh scattering component $I_{ray}(\lambda)$, (ii) the aerosol scattering component $I_{aer}(\lambda)$ and (iii) the component that accounts for multiple reflection of irradiance between the ground and the air $I_{rf}(\lambda)$

$$I_{dif}(\lambda) = I_{ray}(\lambda) + I_{aer}(\lambda) + I_{rf}(\lambda). \tag{6}$$

The Rayleigh scattered component $I_{ray}(\lambda)$ of diffuse part of UV irradiance is calculated as

$$I_{ray}(\lambda) = I_0(\lambda) T_{O_3}(\lambda) T_{SO_2}(\lambda) T_{NO_2}(\lambda) T_{aa}(\lambda) \left(1 - T_{ray}^{0.95}(\lambda)\right) / 2. \tag{7}$$

$T_{O_3}$, $T_{SO_2}$, $T_{NO_2}$, $T_{aer}$ and $T_{ray}$ are $O_3$, $SO_2$, $NO_2$, aerosol and air transmittances that have been defined previously. Transmittance of the aerosol absorption process, $T_{aa}(\lambda)$, is defined in Ref. 30 as

$$T_{aa}(\lambda) = exp\left[-(1 - \omega(\lambda))\tau_a(\lambda)\right], \tag{8}$$

where $\omega(\lambda)$ is the single-scattering albedo, and $\tau_a(\lambda)$ is aerosol optical thickness.

The aerosol-scattered irradiance is calculated as

$$I_{aer}(\lambda) = I_0(\lambda) T_{O_3}(\lambda) T_{SO_2}(\lambda) T_{NO_2}(\lambda) T_{aa}(\lambda) T_{ray}^{1.5}(\lambda) \left[1 - T_{as}(\lambda)\right] D_s(\lambda), \tag{9}$$

where $T_{as}(\lambda)$ is the transmittance for aerosol scattering, such that

$$T_{as}(\lambda) = exp[-\omega(\lambda)\tau_a(\lambda)] \qquad (10)$$

and $D_s(\lambda)$ is the fraction of the scattered flux that is transmitted downwards. The function $D_s(\lambda)$ is dependent on the aerosol asymmetry factor $\delta$ and solar zenith angle $\theta$, according to Bird & Riordan [28] and Justus & Paris [30] as

$$D_s = F_s C_s, \qquad (11)$$

$$F_s = 1 - 0.5 exp[(B_1 + B_2 cos\theta) cos\theta], \qquad (12)$$

$$B_1 = B_3[1.459 + B_3(0.1595 + B_3 \times 0.4129)], \qquad (13)$$

$$B_2 = B_3[0.0783 + B_3(-0.3824 - B_3 \times 0.5874)], \qquad (14)$$

$$B_3 = ln(1-\delta), \qquad (15)$$

$$C_s(\lambda) = (\lambda + 0.55)^{1.8}. \qquad (16)$$

The asymmetry factor is a key optical characteristic of aerosols and it is used from OPAC database [26] for each wavelength and humidity.
Backscattered component of multiple reflections between air and ground is calculated following Bird & Riordan [28] as

$$I_{rf}(\lambda) = \frac{[I_{dir}(\lambda) + I_{ray}(\lambda) + I_{aer}(\lambda)] r_s(\lambda) r_g(\lambda) C_s(\lambda)}{1 - r_s(\lambda) r_g(\lambda)}, \qquad (17)$$

where $r_g(\lambda)$ is ground albedo and $r_s(\lambda)$ is sky reflectivity. Ground albedo is used from Ruggaber et al.[22] while sky reflectivity is calculated by

$$r_s(\lambda) = T'_{O_3}(\lambda) T'_{aa}(\lambda)[0.5(1 - T'_{ray}(\lambda)) + (1 - F'_s(\lambda)) T'_{ray}(\lambda)(1 - T'_{as}(\lambda))], \qquad (18)$$

where the primed transmittance terms are the regular atmospheric transmittance evaluated at optical mass of 1.8. . More details about this model are elaborated in Ref. 5.

### 2.3. Description of the Vojvodina region and time series

The Vojvodina region (Serbia) is situated in the northern part of Serbia and the southern part of the Pannonian lowland (18°51′–21°33′E, 44°37′–46°11′N and 75–641m a.s.l.) (Fig. 2). For the complexity analysis of the UV-B radiation dose time series in this paper we selected the following places: Sombor (SO), Subotica (SU), Novi Sad (NS), Kikinda (KI), Zrenjanin (ZR), Banatski Karlovac (BK) and Sremska Mitrovica (SM) as shown in Fig. 1. The UV-B radiation has a pronounced impact on the human health and some plants in agricultural activities in this region that is the most important food production area in Serbia with surface area of 21,500 km$^2$ and a population of about 2 million people. Monitoring details of the UV-B radiation in the Vojvodina region are given in Ref. 8.

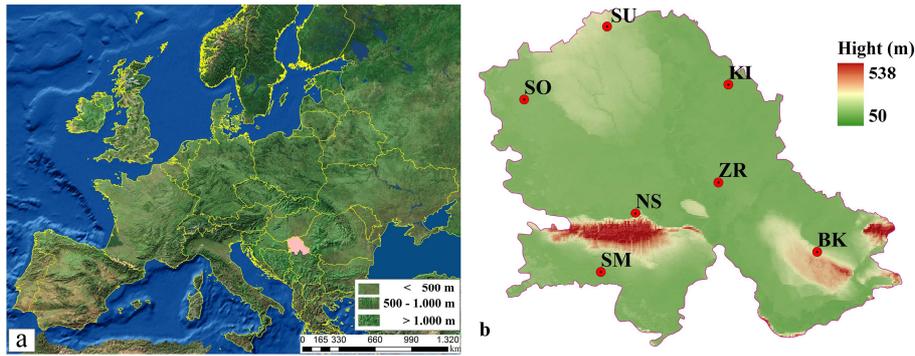

Fig. 2. Location of the Vojvodina region (Serbia) in the Europe (a) and places used in study (b); the places are: Sombor (SO), Subotica (SU), Novi Sad (NS), Kikinda (KI), Zrenjanin (ZR), Banatski Karlovac (BK) and Sremska Mitrovica (SM).

We have formed the corresponding time series combining three sources because of the lack of measurement places for the UV radiation in the Vojvodina region. We have included: (i) values measured in Novi Sad (45.33° N, 19.85° E, 84 m a.s.l.) measured by the broadband Yankee UVB-1 biometer, (ii) values computed by a parametric numerical model and (iii) values computed by a empirical formula based on linear correlation between the daily dose of the UV-B ($UVB_d$) and the daily sum of the global solar radiation ($G_d$) in MJ m$^{-2}$.[9] The empirical formula, which is derived on the basis of relationship between daily values of $UVB_d$ (measured UV-B data and corresponding calibration factors) and $G_d$ (computed via an empirical formulae) for the period April 2003 - December 2009 in Novi Sad (correlation coefficient R=0).

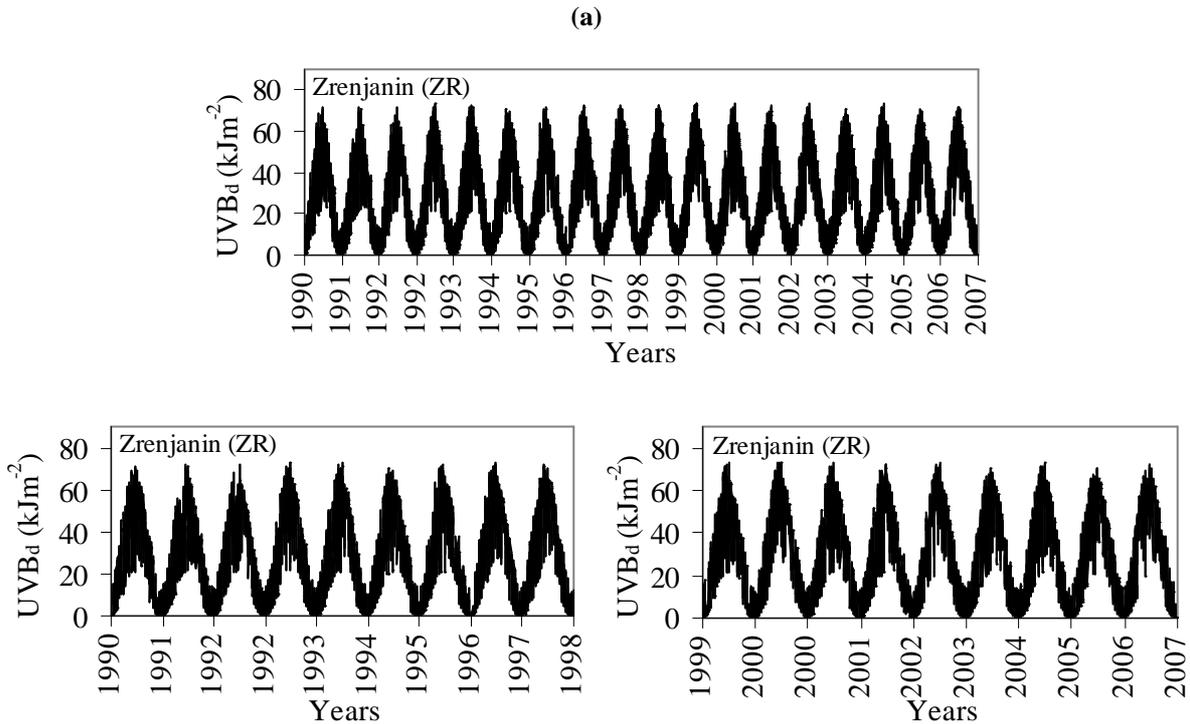

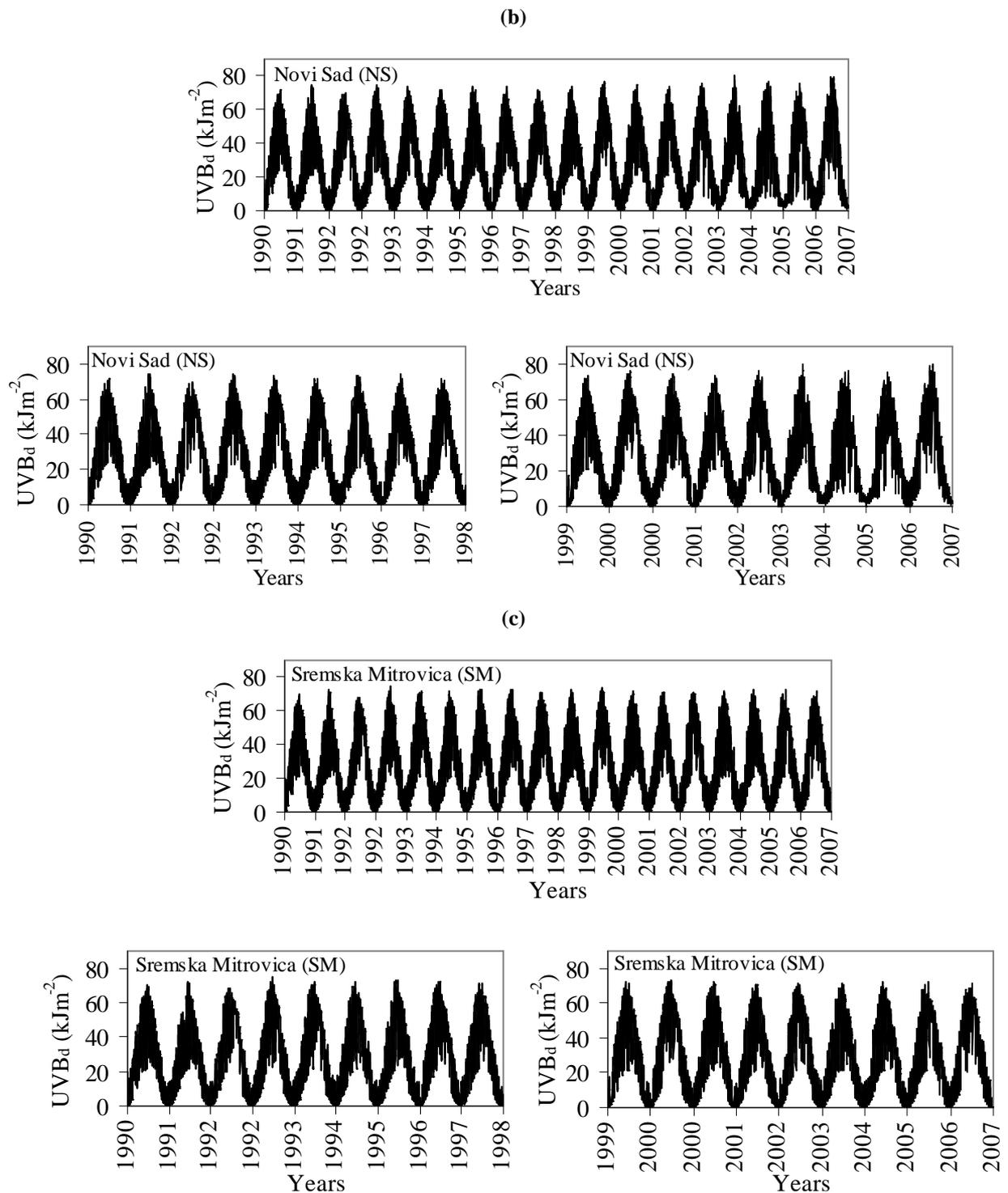

Fig. 3. The UV-B radiation dose time series (1990-2007, 1990-1998, 1999-2007 year) for three places in the Vojvodina region (Serbia) analyzed for this paper.

## 2.4. Methodological details

Using the calculation procedure described in the subsection 2.1 and Appendices A and B, we have computed the KL, KLM and SE values for the seven UV-B radiation dose time series (Fig. 3). The calculations are carried out for the entire time interval 1990-2007 and for two subintervals covering this period: (a) 1990-1998 and (b) 1999-2007. They are calculated by decomposing a sequence into a production history, but in different ways. The sequence decomposition occurs at points where the eigen function increases in value from the previous one. In this case, these points are the locations where an extra symbol causes an increase in the accumulated vocabulary. The exhaustive complexity calculation is based on finding extensions to a sequence, which are not reproducible from that sequence, using a recursive symbol-copying procedure. Exhaustive complexity can be considered a lower limit of the complexity measurement approach proposed in the LZA, and primitive complexity an upper limit [10, 31]. All complexity measures are sensitive to the length of time series, $N$. For the SE (Appendix B), there exists a recommendation for use $N$ that is larger than 200.[32] For the time interval 1990-2007 and two subintervals (1990-1998 and 1999-2007), the length of time series was $N$ = 6574, 3287 and 3287, respectively. Let us note that Hu et al.[33] derived analytic expression for $C_k$ (notation in Appendix A) in the KL, for regular and random sequences. In addition they showed that the shorter length of the time series, the larger $C_k$ value and correspondingly the complexity for a random sequence can be considerably larger than 1. The SE is sensitive on input parameters: embedding dimension ($m$), tolerance ($r$) and time delay ($\tau$). In this paper it was calculated for UV-B radiation dose time series with the following values of parameters: $m$ =2, $r$ =0.2 and $\tau$ =1.

## 3. RESULTS AND COMMENTS

We have computed the KL, KLM and SE values for the UV-B radiation dose time series of seven places. The computations are carried out for the entire time interval 1990-2007. The results are listed in Table 1. It is seen from this table that for five places (Sombor, Subotica, Novi Sad, Kikinda and Zrenjanin) their KL values are close each to other (0.492, 0.498, 0.492, 0.496 and 0.498), i.e., they are practically the same. However, in contrast to these places for Sremska Mitrovica and Banatski Karlovac have higher values of the KL (0.523 and 0.515). Since Sremska Mitrovica is close to Fruška Gora Mountain while Banatski Karlovac is located in a hilly region (Fig. 3), the increase of the complexity in those places can be attributed to enhanced UV-B radiation dose caused by the multiple scattering effects.[34,35] Following this reason it could be expected that Novi Sad, which is also in the vicinity of the Fruška Gora mountain, has the higher level of the complexity. However, this place is highly urbanized with more emission sources in comparison with Sremska Mitrovica, thus the urban air pollution reduces the amount of UV-B radiation reaching the ground. Namely, according to Bais et al.[36] the surface UV-B radiation at locations near the emission sources of $O_3$, $SO_2$ or $NO_2$ in the lower troposphere is attenuated by up to 20%. In result, the complexity of the UV-B radiation dose decreases. Note, if a process is less complex then it has a KL value close to zero, whereas a process with highest complexity will have the KL close to one. In addition, the KL measure can be also considered as a measure of randomness. Thus, a value of the KL near zero is associated with a simple deterministic process, while a value close to one is associated with a stochastic process. If we look at the KLM values we reach the same conclusions. To our knowledge, the KL and KLM measures has not been used for analyzing the complexity of the UV-B radiation dose time series.

In our analysis we have employed another complexity measure, i.e., the SE, which is not often in the analysis of the complexity of geophysical time series unlike to approximate entropy.[37] Such analysis was done by Shuangcheng et al.[38] in measurement of climate complexity using daily temperature time series. The calculated values of the SE are also listed in Table 1. Those values, which are close each to other, indicate on a similar behaviour of UV-B radiation dose time series for the entire time interval 1990-2007 and all places , i.e. their lower irregularity.

| Place | Measure | 1990-2007 | 1990-1998 | 1999-2007 |
|---|---|---|---|---|
| SO (45°47'N, 19°05'E) | KL | 0.492 | 0.519 | 0.505 |
|  | KLM | 0.511 | 0.526 | 0.522 |
|  | SE | 1.206 | 1.203 | 1.176 |
| SU (46°06'N, 19°46'E) | KL | 0.498 | 0.498 | 0.489 |
|  | KLM | 0.512 | 0.522 | 0.522 |
|  | SE | 1.245 | 1.217 | 1.202 |
| NS (45°15'N, 19°51'E) | KL | 0.492 | 0.530 | 0.498 |
|  | KLM | 0.513 | 0.547 | 0.512 |
|  | SE | 1.223 | 1.262 | 1.174 |
| KI (45°51'N, 20°28'E) | KL | 0.496 | 0.526 | 0.501 |
|  | KLM | 0.509 | 0.533 | 0.522 |
|  | SE | 1.238 | 1.216 | 1.146 |
| ZR (45°24'N, 20°21'E) | KL | 0.498 | 0.544 | 0.487 |
|  | KLM | 0.527 | 0.565 | 0.526 |
|  | SE | 1.238 | 1.252 | 1.233 |
| SM (44°58'N, 19°38'E) | KL | 0.523 | 0.551 | 0.508 |
|  | KLM | 0.536 | 0.572 | 0.533 |
|  | SE | 1.252 | 1.234 | 1.178 |
| BK (45°03'N, 21°02'E) | KL | 0.515 | 0.530 | 0.530 |
|  | KLM | 0.532 | 0.558 | 0.530 |
|  | SE | 1.191 | 1.246 | 1.243 |

**Table 1.** Kolmogorov complexity (KL), Kolmogorov complexity spectrum highest value (KLM) and sample entropy (SE) values for the UV-B radiation dose time series of seven places in the Vojvodina region (Serbia) for the period 1990-2007, and the subintervals: (a) 1990-1998 and (b) 1999-2007. In computing the entropy we have used the following sets of parameters ($m$ =2, $r$ =0.2 and $\tau$ =1).

We have also divided the period 1990-2007 into two subintervals: (a) 1990–1998 and, (b) 1999-2007, and calculated the KL and SE values for the various time series in each of these subintervals. These intervals were chosen because it was expected a change in the complexity of the UV-B radiation dose after 1999 in the Vojvodina region when occurred: (i) a large increase of air and soil pollution (period after the civil war, air military activities, increase of the industrial activities frozen in the previous period, higher traffic frequency etc.) and (ii) an increase of cloudiness due to climate change[39] and corresponding influence on UV radiation dose.[40,41] Let us note that the KL complexity of different kind of biomedical, hydrological and physical time series may be lost due to different reasons that come from reducing the functionality of some system segments represented by those series. For example, Gomez & Hornero [42] using entropy and complexity analyses of Alzheimer's disease (AD) have showed that the complexity reduction seems to be associated with the deficiencies in information processing suffered by AD patients. And another example from the river flow time series analysis by Orr and Carling [43] points out that the complexity loss may be attributed to the extent of human intervention involving land and crop use, urbanization, commercial navigation and other activity. Thus, decrease of the KL complexity of some process represented by a time series is an indicator of a simplification of that process caused by some crucial agent.

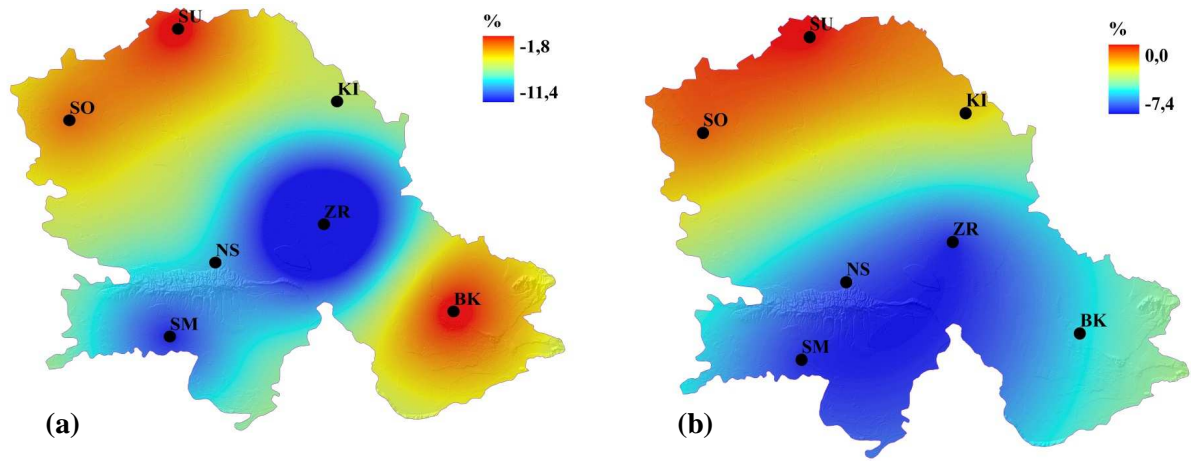

Fig. 4. Relative change of the KL (a) and KLM (b) from the period 1990-1998 comparing to the period 1999-2007 for places in the Vojvodina region (Serbia). Abbreviations are the same as in Fig. 2.

It is found that during 1999-2007, there is a decrease in complexity in all places (Sombor - 0.505; Subotica - 0.489; Novi Sad - 0.498; Kikinda - 0.501; Zrenjanin - 0.487; Sremska Mitrovica - 0.508 and Banatski Karlovac - 0.530) in comparison to the period 1990-1998 (Sombor - 0.519; Subotica - 0.498; Novi Sad - 0.530; Kikinda - 0.526; Zrenjanin - 0.544; Sremska Mitrovica - 0.551 and Banatski Karlovac - 0.539) as it presented in Table 1. These differences are visualized in Fig. 4. It shows relative change of the KL (Fig. 4a) and KLM (Fig. 4b) from the period 1990-1998 comparing to the period 1999-2007 for the seven places. From Fig. 4 it is seen that the central and south western parts of the Vojvodina region have the largest decline of the KL (Fig. 4a) and KLM (Fig. 4b) complexities. In other parts that decline is much lower. Among places with the large decline of both complexities, Zrenjanin stands out with the largest one. It is result of a very large concentration of $SO_2$ and particles in this place that come from the mentioned human activities. Namely, $SO_2$ absorbs radiation in the UV-B part of the spectrum, remarkably affecting the reduction of the UV-B radiation through sulphate aerosols. It is estimated that in the industrialized countries on the northern hemisphere sulphate aerosols can reduce the UV-B radiation for 5-18% [44]. Figure 5 depicts the KL complexity spectrum of the normalized UV-B radiation dose for three places (Zrenjanin, Novi Sad and Sremska Mitrovica). From this it is seen that, for all places, the highest differences in spectra of complexity (period 1990-1998 versus period 1999-2007) are in the interval (0.3, 0.5) of the normalized UV-B radiation doses. Finally, note that there exists a set of programs in nonlinear time series analysis, for determining the dynamics of the UV-B radiation, which can support the analysis through measures we suggested in this paper. [45,46,47,48]

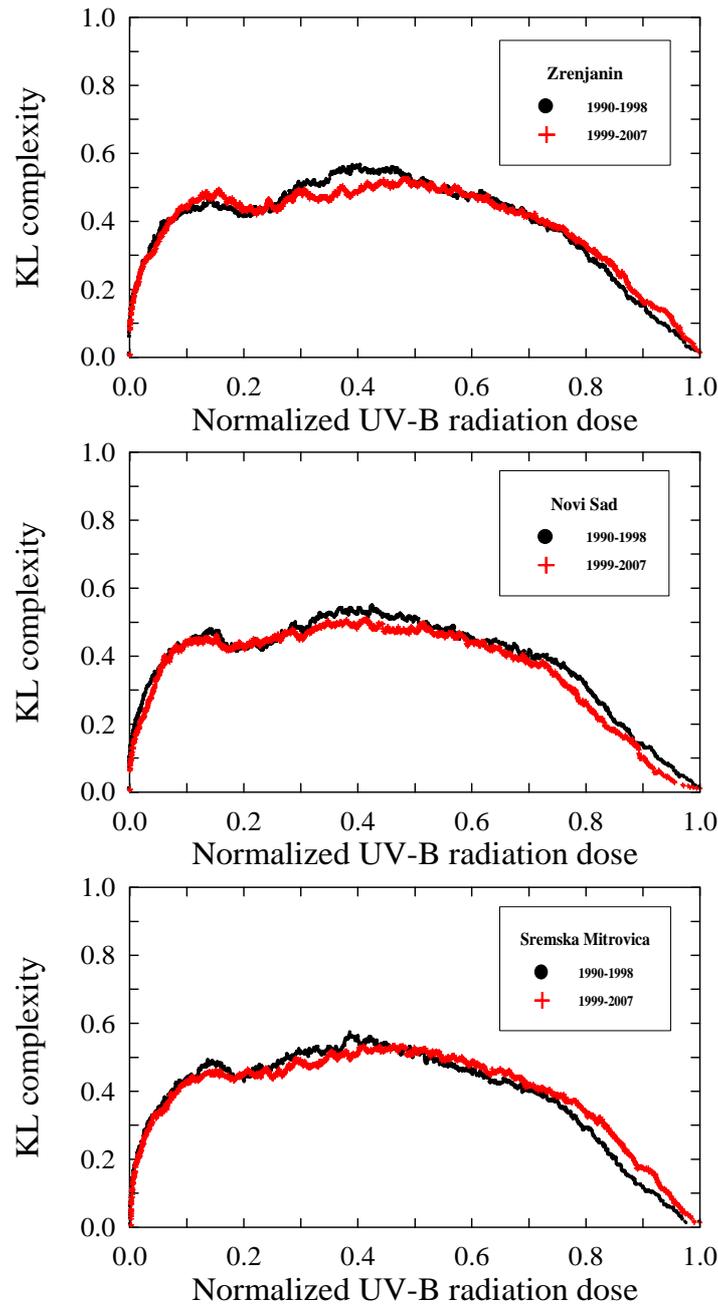

Fig. 5. The Kolmogorov complexity spectrum of the UV-B radiation dose time series for three places in the Vojvodina region (Serbia). On $x$ axis are depicted the values of the time series normalized as $x_i = (X_i - X_{min})/(X_{max} - X_{min})$, where $\{X_i\}$ is the time series of the UV-B radiation dose obtained by procedures described in subsection 2.3 and $X_{max} = max\{X_i\}$ and $X_{min} = min\{X_i\}$.

## 4. CONCLUDING REMARKS

In this paper we have analyzed UV-B radiation to assess the complexity in UV-B radiation dose in the Vojvodina region (Serbia) for the period 1990-2007. We have defined the Kolmogorov complexity spectrum and have introduced the Kolmogorov complexity spectrum highest value (KLM). We have examined the daily dose of UV-B radiation time series from seven places (Sombor, Subotica, Novi Sad, Kikinda, Zrenjanin, Sremska Mitrovica and Banatski Karlovac) and calculated the KL, KLM and SE values for each time series.

The UV-B radiation dose time series were established on the basis of their daily sum using: (i) measured data, (ii) data calculated via a derived empirical formula and (iii) data obtained by a non-parametric UV-B radiation model. According to all computed measures it is found that during 1999–2007, there is a decrease in complexity in all places in comparison to the period 1990–1998. This complexity loss may be attributed to (i) the increased human intervention in the post civil war period causing increase of the air pollution after 1999 and (ii) the increased cloudiness due to climate changes.

**ACKNOWLEDGMENTS**

The research presented in this paper was realized as a part of the project "Studying climate change and its influence on the environment: impacts, adaptation and mitigation" (No. III 43007) supported by the Ministry of Education and Science of the Republic of Serbia within the framework of integrated and interdisciplinary research over the period 2011–2014. The authors are grateful to the Provincial Secretariat for Science and Technological Development of Vojvodina for the support under the project "Climate projections for the Vojvodina region up to 2030 using a regional climate model" funded by (No. 114-451-2151/2011-01).

**Appendix A.**

**Description of the LZA for computing the KL complexity**

The KL complexity of a time series $\{x_i\}$, $i = 1, 2, 3, 4, ..., N$ by the LZA, can be summarized as follows. Step 1: Encode the time series by constructing a sequence $s$ consisting of the characters 0 and 1 written as $\{s(i)\}$, i=1,2,3,4,…,N, according to the rule

$$s(i) = \begin{cases} 0 & x_i < x_* \\ 1 & x_i \geq x_* \end{cases}. \tag{A.1}$$

Here $x_*$ is a threshold, that should be properly chosen. The mean value of the time series has often been used as the threshold.[49] Depending on the application, other encoding schemes are also available.[50] Step 2: Calculate the complexity counter $C(N)$, which is defined as the minimum number of distinct patterns contained in a given character sequence;[51] $c(N)$ is a function of the length of the sequence N. The value of $c(N)$ is approaching an ultimate value b(N) as N approaching infinite, i.e.

$$c(N) = O(b(N)), \quad b(N) = \frac{N}{\log_2 N}. \tag{A.2}$$

Step 3: Calculate the normalized complexity measure $C_k(N)$, which is defined as

$$C_k(N) = \frac{c(N)}{b(N)} = c(N) \frac{\log_2 N}{N}. \tag{A.3}$$

The $C_k(N)$ is a parameter to represent the information quantity contained in a time series, and it is to be a 0 for a periodic or regular time series and to be a 1 for a random time series, if N is large enough. For a non-linear time series, $C_k(N)$ is to be between 0 and 1.

**Appendix B.**

**Calculation of the sample entropy**

This is a measure quantifying regularity and complexity, it is believed to be an effective analyzing method of diverse settings that include both deterministic chaotic and stochastic processes, particularly operative in the analysis of physiological, sound, climate and environmental interface signals that involve relatively small amount of data.[14, 52-54] The threshold factor or filter $r$ is an important parameter. In principle, with an infinite amount of data, it should approach zero. With finite amounts of data, or with measurement noise, $r$ value typically varies between 10 and 20 percent of the time series standard deviation. To calculate the from a time series, $X = (x_1, x_2, ..., x_N)$, , one should follow these steps.[14]:

(1) Form a set of vectors $X_1^m, X_2^m, ..., X_{N-m+1}^m$ defined by $X_i^m = (x_i, x_{i+1}, ..., x_{i+m-1})$, $i = 1, ..., N-m+1$;

(2) The distance between $X_i^m$ and $X_j^m$, $d[X_i^m, X_j^m]$ is the maximal absolute difference between their respective scalar components: $d[X_i^m, X_j^m] = \max_{k \in [0, m-1]} |x_{i+k} - x_{j+k}|$;

(3) For a given $X_i^m$, count the number of $j$ $(1 \leq j \leq N-m, j \neq i)$, denoted as $B_i$, such that $d[X_i^m, X_j^m] \leq r$. Then, for $1 \leq i \leq N-m$, $B_i^m(r) = B_i / (N-m-1)$;

(4) Define $B_i^m(r)$ as: $B^m(r) = \{\sum_{i=1}^{N-m} B_i^m(r)\} / (N-m)$;

(5) Similarly, calculate $A_i^m(r)$ as $1/(N-m-1)$ times the number of $j$ $(1 \leq j \leq N-m, j \neq i)$, such that the distance between $X_j^{m+1}$ and $X_i^{m+1}$ is less than or equal to $r$. Set $A^m(r)$ as: $A^m(r) = \{\sum_{i=1}^{N-m} A_i^m(r)\} / (N-m)$. Thus, $B^m(r)$ is the probability that two sequences will match for $m$ points, whereas $A^m(r)$ is the probability that two sequences will match $m+1$ points;

(6) Finally, define: $SampEn(m, r) = \lim_{N \to \infty} \{-\ln[A^m(r) / B^m(r)]\}$ which is estimated by the statistic:

$$SampEn(m, r, N) = -\ln \frac{A^m(r)}{B^m(r)}.$$